\documentclass[page-classic]{epl2arXiv} 
\usepackage{amssymb}
\usepackage{graphics}
\usepackage{subfig}
\usepackage{epsfig}
\usepackage{color}

\title{A principle for ideal torus knots}

\author{Kasper W Olsen\inst{1} \and Jakob Bohr\inst{1}}
\shortauthor{Kasper W Olsen and Jakob Bohr}

\institute{                    
DTU Nanotech,
Building 345E, 
\O rsteds Plads,
Technical University of Denmark,
DK-2800 Kongens Lyngby, Denmark
}

\pacs{02.60.Pn}{Numerical optimization}
\pacs{02.10.Kn}{Knot theory}
\pacs{87.15.-v}{Biomolecules}

\abstract{
We study simple, knotted and linked torus windings that are made of tubes of finite thickness. Knots which have the shortest rope length are often denoted ideal structures. Conventionally, the ideal structure are found by rope shortening routines. 
It is shown that alternatively they can be directly determined as maximally rotated structures. In many cases these structures are also zero-twist structures i.e. structures that neither rotate one or the other way under strain. We use this principle to implement rapid numerical calculations of the ideal structures and subsequently quantify them by their aspect ratio.
The results are compared with the aspect ratios of biological torus molecules.
}

\begin{document}

\maketitle

\section{Introduction}
Knots which are made with as little amount of rope as possible have been denoted {\it ideal knots} \cite{katritch1996,knots1998}. Physical knots are made of circular tubes of finite thickness. Therefore, they are of interest in modeling real physical systems with this topology, e.g. polymers.
In the mathematics literature, the rope length $L(K)$  
has a precise meaning -- it is the length of the curve defining the knot divided by its thickness. 
In general, it is not known what the minimal rope length is, and which is the corresponding knot geometry. 
The problem has only been solved for the trivial knot, it has rope length $2\pi$. For any other knot a lower bound has been found for the rope length, 
$L(K) \geq 15.66$ \cite{denne2006}, but the knots have only been optimized on a case-by-case basis using different algorithms. 

In many minimalization problems one does not have a way of determining if an optimum state has been reached. Examples are common algorithms for NP complete problems. This is also the case for the  relatively simple problem of shortening a thick knot for which successful algorithms have been proposed \cite{pieranski1998,ashton2011}, and the energy minimalizations \cite{ohara1991,ohara2011}.
For knots, we present a zero-twist analysis that can reveal if they are absolutely shortened or not. Further, in many cases a corresponding algorithm can be used to compute the optimal knots.

In this short note, a small step towards finding a general principle for characterizing ideal knots is taken. A method for determination of ideal torus knots, i.e. knots restricted to having a toroidal geometry, is suggested using our principle of maximally rotated structures introduced for ropes\cite{bohr2011a,olsen2012}. 
For other discussions of rope-like structures, see \cite{przybyl2001,neukirch2002,bruss2012}. Although tube models have been compared to the folding patterns of proteins \cite{banavar2003a,banavar2003b}, the interest in knots of toroidal geometry have traditionally been mostly mathematical.  Here, we complement these structures with comparisons of the optimal knots with the highly symmetrical biological torus molecules. 

\begin{figure}[h]\centering
\includegraphics[width=5.1cm]{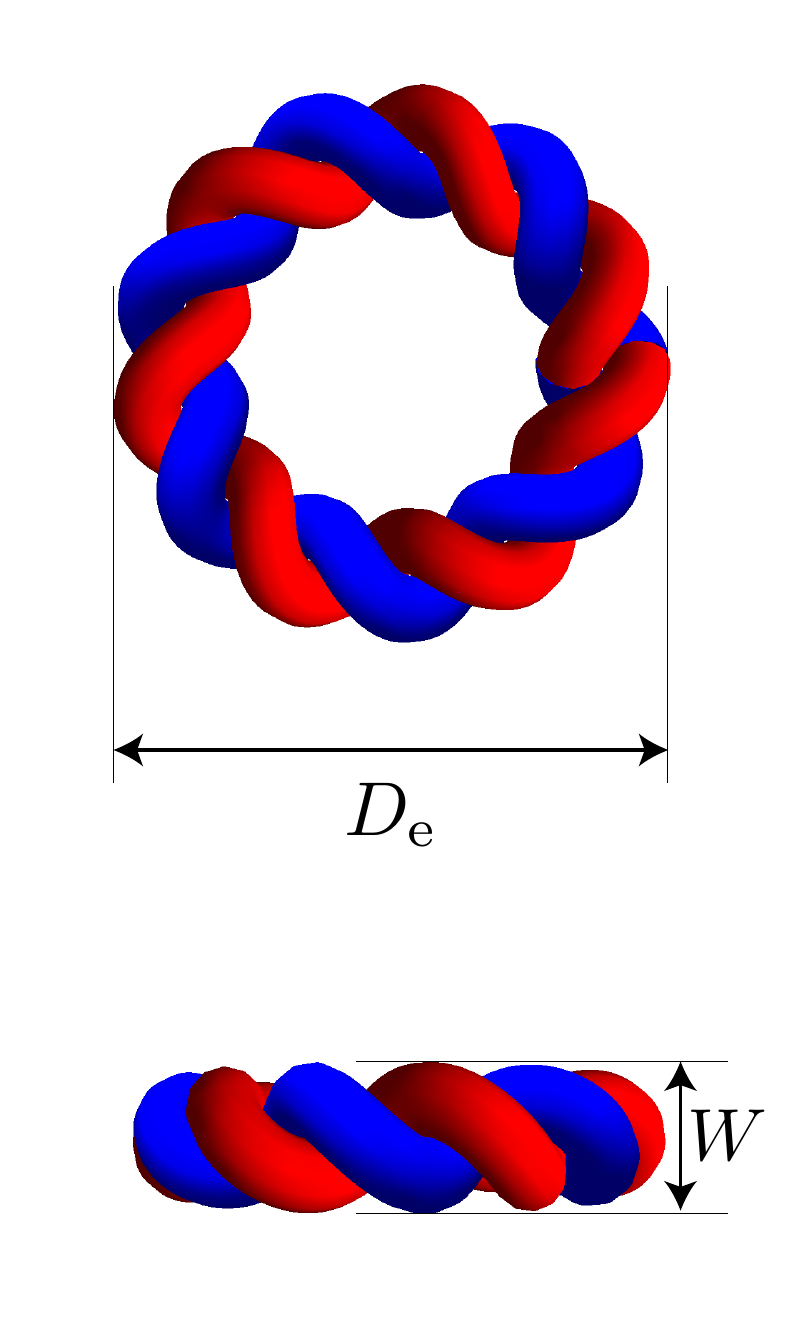}
\caption{Tube model of torus molecule viewed perpendicular (top) and parallel (bottom) to the plane of the torus. The tube diameter is $D$. Indicated in the figure is the exterior diameter, $D_{\rm e}=2(R+a)+D$,  of the torus molecule, and its width, $W=2a+D$. The specific knot depicted here is a $T(13,2)$ knot.}
\label{tubemodel}
\end{figure}

The conditions on how tubes can produce closed loops, thereby forming torus knots and links are straightforward.
Specifically, a $(p,q)$-torus knot denoted $T(p,q)$ is obtained by looping a tube on a torus $p$ times with $q$ revolutions before joining its ends.
A torus knot is topologically trivial, i.e. a unknotted loop, when either $p$ or $q$ is equal to 1. The simplest non-trivial example is the $T(3,2)$ knot, which topologically is a trefoil knot.
A parametric equation for the $T(p,q)$ knot is
\begin{eqnarray}
\label{pqknot}
x &=& (R+a\cos(p t))\cos(qt) \nonumber\\
y&=&(R+a\cos(p t))\sin(qt)\\
z&=&a\sin(pt) \nonumber
\end{eqnarray}
for $0\leq t < 2\pi$. In this note, the torus knots are wrapped on a torus, where $a$ is the helix radius and $R$ the torus radius; $D$ will denote the thickness of the knot, see fig. \ref{tubemodel}.

The literature on the study of ideal knots is well established \cite{knots1998}.
Yet the problem of how to minimize the length of knots, and other questions have not been answered. E.g. what is the length, $L(3,2)$, of the simplest non-trivial ideal knot?
Earlier, Przyby{\l} {\it et al.} have found $L(3,2) = 17.0883$ for the ideal torus knot \cite{pieranski2001}. 
By deviating from the class of torus knots which have a simple and fixed symmetry, Piera\'{n}ski and Przyby{\l} have suggested
an ideal $(3,2)$ knot using the SONO algorithm; their best estimate is $L(3,2)=16.38$ \cite{pieranski2001}. Baranska, Przyby{\l}, and Piera\'{n}ski investigated the 
suggested ideal (3,2) knot, including its global curvature and torsion, further in \cite{baranska2008}.
Ideal trefoil knots have been investigated by Baranska {\it et al.} \cite{baranska2008,baranska2004}, and also discussed in a paper by Snir and Kamien \cite{snir2007}. Ideal knots have also been studied by Gerlach \cite{gerlach2010}. 
An overview of rope lengths for knots and links is Cantarella {\it et al.} \cite{cantarella2002}. According to Gonzales and Maddocks, the thickness of a knot can also  be defined in terms of a global radius of curvature \cite{gonzales1999}. Tight open knots have been studied by Piera\'{n}ski {\it et al.} \cite{pieranski2001a}; and polygonal knots give an upper bound on the rope length of some ideal knots \cite{pieranski2004}. Ashton {\it et al.} \cite{ashton2011} have computed the rope length and self-contact points of 379 tight packed knots and links by using a gradient descent method. The trefoil knot has also appeared locally on long molecular chains studied by Monte Carlo algorithms \cite{metzler2002}.

\section{Method}

First,  consider $T(p,2)$ knots and links. As knots and links are closed curves, one can build the $(2m,2)$ torus links and $(2m+1,2)$ torus knots from joining the ends of a {\it circular double helix}.
The geometry of the typical circular double helix has been investigated in \cite{olsen2012}.
The double helix that makes a closed loop after one revolution becomes the $(2m,2)$ torus link.
The period, $T$, of the parametrization must obey $T=2m \pi$, where $R/h=m \in {\bf N}$; here $R$ is the torus radius and $h$ is the reduced helical pitch, $h=H/2\pi$; the helical pitch is denoted by $H$.
For the double helix, where the strands make a closed loop after two revolutions the period obeys $T=(2m+1)\pi$, and $R/h=m+1/2$, $m\in {\bf N}$. The knot is a $(p,2)=(2m+1,2)$ torus knots. For the double helix, this completes the set of possible torus knots and links. For simplicity, we only discuss how to find ideal torus knots.

Here, we describe two different ways of optimizing such torus structures. First, with respect to incremental twist, $f_\Theta$, and then with respect to volume fraction $f_{\rm V}$.
Mathematically, an {\it ideal torus knot} is one with a minimal rope length under the condition that it is confined to a torus and has a trigonometrical parametrization equivalent to eqs. (\ref{pqknot}). Imagine that the radius of the torus is very large, in this case
the knot is tied on what is essentially a straight cylinder. 
The knot is therefore locally a two-stranded rope.
The zero-twist structure of a rope has the property that it is maximally twisted, and this secures that the rope-length of the strands are minimized \cite{bohr2011a}. Non-maximally twisted structures can reduce their rope length,
per $2\pi$ progression of the rope, upon further twisting. Similarly, the ideal torus knots are the maximally twisted double helices with closed circular geometry. 
Therefore, in our notation the ideal torus knot of length $L$ is the knot that maximizes $f_\Theta=D\Theta/L$, where $\Theta$ is the twist angle. The function $f_\Theta$ is called the incremental twist in ref. \cite{olsen2012}, and its maxima are helices that neither rotate one or the other way under strain. The incremental twist, $f_\Theta$, for a toroidal helix reads,
\begin{equation}
f_\Theta = \frac{D}{a}\left( 
\frac{1}{2\pi}\int_0^{2\pi}\sqrt{1+\frac{H}{2a}(1-\frac{a}{R}\cos\phi)^2}\, d\phi
\right)\, .
\end{equation}
A way to find ideal knots is the following. Take as special case the $T(p,2)$ knot, where we have $R=(p/2)h$. Calculate the incremental twist $f_\Theta(D/R,H/D)$. Then we have

\begin{equation}
\frac{df_\Theta (R/D,H/D)}{d(H/D)} = \frac{\partial f_\Theta}{\partial (R/D)}\frac{\partial (R/D)}{\partial (H/D)}
+\frac{\partial f_\Theta}{\partial (H/D)}
\end{equation}
The zero-twist condition is
\begin{equation}
\frac{\partial f_\Theta}{\partial (H/D)} = 0
\end{equation}
The condition for ideal knot (with extrema) is
\begin{equation}
\frac{\partial f_\Theta}{\partial (R/D)}\frac{p}{4\pi}
+\frac{\partial f_\Theta}{\partial (H/D)}=0
\end{equation}
Here we have calculated along the line $R/D=(p/2)h/D=(p/4\pi)H/D$, since this is the condition for the strand to make a closed loop. Only under certain conditions of proper structures is the ideal knot criterion the same as the zero-twist criterion.

\begin{figure}[h]\centering
\includegraphics[width=7.1cm]{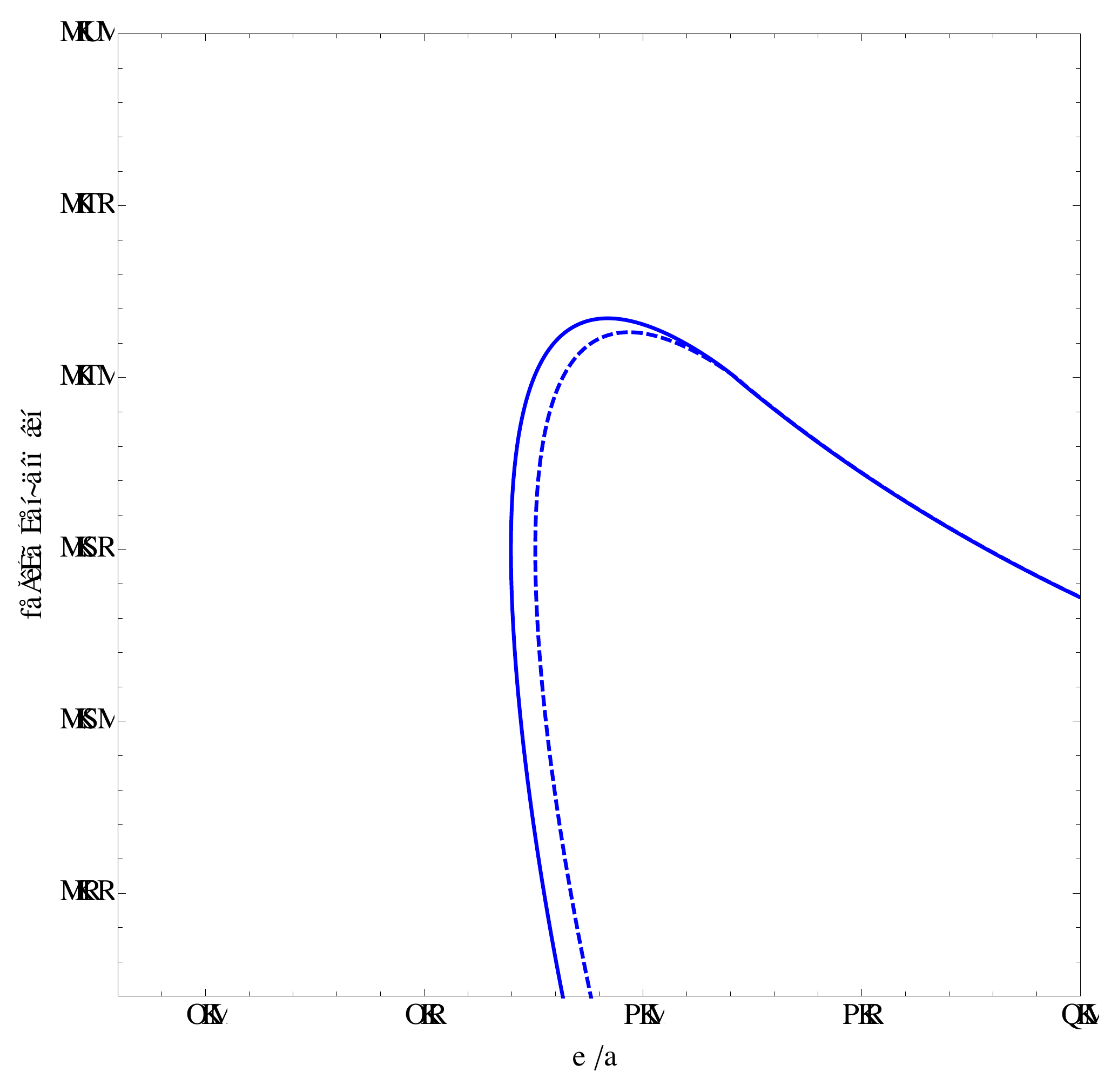}
\caption{Incremental twist, $f_\Theta$, as a function of dimensionless pitch, $H/D$, for the class of torus knots $T(19,2)$ (solid line) and $T(17,2)$ (dashed line) with inter-strand contact. The maximum of $f_\Theta$ at the apex is at $H/D=2.98489$ and $H/D=3.01919$, respectively. The corresponding ideal torus knot has this dimensionless pitch respectively.}
\label{isoknot}
\end{figure}

As the ideal torus knots maximize $f_\Theta$, they can be found in the following simple way. 
Figure \ref{isoknot} shows a plot of $f_\Theta$ as a function of the dimensionless pitch, $H/D$, for two examples of knot classes: The "isoknot" lines in fig. \ref{isoknot} correspond to the family of $T(17,2)$ and $T(19,2)$ torus knots with inter-strand contact.
Among these knots we find the one which maximize $f_\Theta$ at the apex of the isoknot line.

Another way of optimizing the torus structures is by maximizing the volume fraction, i.e. finding close-packed structures. The volume fraction, $f_{\rm V}$, is here defined as the local volume of the two tube compared to the volume of an enclosing bent cylinder. For a toroidal double helix, 

\begin{equation}
f_{\rm V} =  \left( \frac{2a}{D} +1\right)^{-2} \frac{1}{\pi} \int^{2\pi}_{0}\sqrt{(\frac{2\pi a}{H})^2 +(1-\frac{a}{R} \cos \phi)^2} \, d\phi\, .
\end{equation}

Having discussed $T(p,2)$ knots and links, we will mention the simpler case of taking $q=1$, i.e. the $T(p,1)$ knot. As said, this is actually an unknotted geometry that is topologically trivial but geometrically it is a slinky which makes $p$ loops around the torus. This circular helix has been studied in \cite{olsen2012}, where the conditions for the tube to be in contact were solved. 
The $T(p,1)$ geometry can be used to model biological molecules with toroidal geometry and $p$-fold symmetry. 

\section{Results}

Detailed results for the knots and links can be found in the Appendix. In Table~\ref{tab:1} numbers for $T(p,2)$ knots are listed. The estimate for the rope length of the ideal torus trefoil is $L(3,2) = 17.088$. In Table~\ref{tab:2} the characteristics of the $T(p,1)$ optimized torus knots are given.
The ideal $T(p,q)$ (with $q\geq 3$) torus knots and links can be found from the closed circular triple helix by a simple generalization of the methods used in the previous section. 

\begin{figure}[h]\centering
\includegraphics[width=8.1cm]{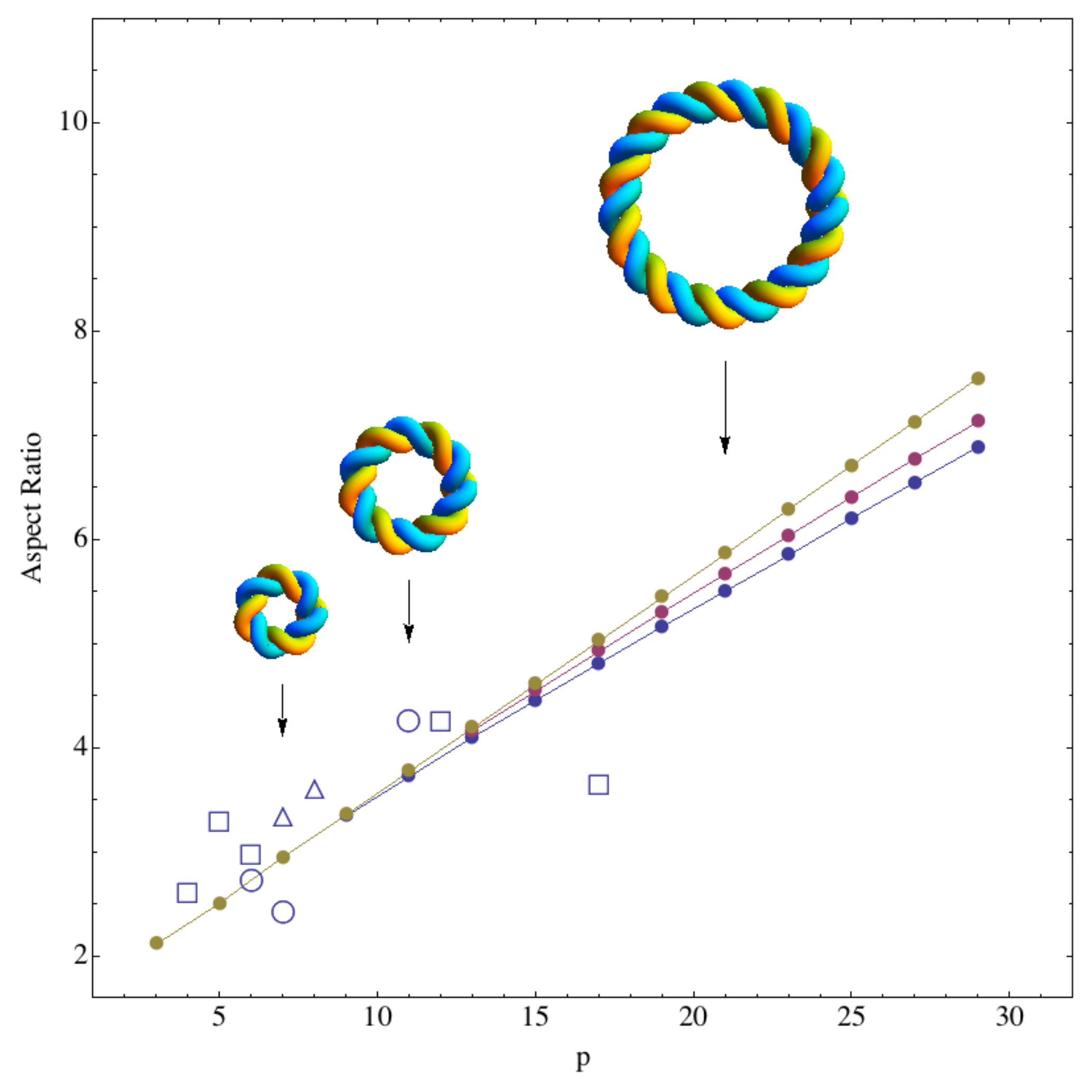}
\caption{Plot of aspect ratio $A= D_{\rm e}/W$ as a function of the number of loops, $p$, for torus knots $T(p,2)$ that are maximally twisted (top) and close-packed (middle) curve. The three depicted knots are from left to right, $T_{\rm ZT}(7,2)$, $T_{\rm ZT}(11,2)$ and $T_{\rm ZT}(21,2)$ corresponding to the yellow point (top curve) below the arrow. The bottom curve is the aspect ratio for the torus slinky $T(p,1)$ which is close-packed. The three symbols, $\square, \bigcirc$, and $\bigtriangleup$, identify ten torus molecules where the aspect ratio has been calculated from molecular structures (Table \ref{tab:3}). Here, the symbol $\square$ is used for viruses, $\bigcirc$ for protein complexes and $\bigtriangleup$ for chaperones.}
\label{aspectratio}
\end{figure}

Figure \ref{aspectratio} is a plot of the aspect ratio $A$ (outer diameter to width) of torus geometries with increasing number of loops, $p$. A fat torus has aspect ratio equal to two. The upper line is the $T_{\rm ZT}(p,2)$ knot  optimized with respect to twist, and the curve below is $T_{\rm CP}(p,2)$ knot optimized by volume fraction. The bottom curve is the $T_{\rm CP}(p,1)$ knots that are optimized with respect to volume fraction. The plot points are measurements on ten different biological torus molecules (viruses, chaperones and protein complexes) using all-atom van der Waals models, see Table~\ref{tab:3}. 

It is seen that within these optimizations, the theoretical aspect ratio is almost a linear function of the number of loops, $p$. Further, to a first approximation the aspect ratio of our measured torus molecules follow the linear trend.


\section{Discussion}

It is unknown which principles are effective in shortening of knots. And if such principles can lead to a unique solution.  The presented zero-twist analysis gives an idea of how to proceed. For the limited case of torus knots, the suggested principle is a direct way of absolute shortening.
The study of closed circular helix structures that are maximally twisted turns out to provide a 
simple analysis of the lower bounds on rope lengths. Data on $T(p,1)$ and $T(p,2)$ knots is included, and it could straightforwardly be extended to larger classes of torus knots. Further, the computation of the rope length of the $T(3,2)$ torus knot gives a value which agrees with previous results in the literature. 

By looking at the experimentally determined aspect ratios of torus molecules, one can even speculate, that  the principle of maximal twist might play a role in determining what the geometry is of biological torus molecules that are formed naturally on the nano-meter scale. In any case, it is remarkable how symmetric Nature's torus molecules can be.

\begin{acknowledgments}
We thank P. Piera\'{n}ski and S. Przyby{\l} for correspondence about ideal knots. This work is supported by the Villum Foundation. 
\end{acknowledgments}

\newpage

\begin{table}
\caption{Ideal $T(p,2)$ torus knots found along the isoknot curves. First column is the knot type, following columns are geometric quantities that characterize the knot. A torus knot is uniquely specified by its value of either $a/R$ and $2a/D$, or by $D/R$ and $H/D$. The fourth column is the rope length, $L(p,2)$ of the knot computed using the criterion of maximal $f_\Theta$. The two last columns are the aspect ratio of knots computed using maximal twist and maximal volume fraction, respectively (for the first three structures, these conditions are identical).}
\label{tab:1}       
\begin{tabular}{cccccc}
\hline\noalign{\smallskip}
Type  & $a/R$  & $H/D$  & $L(p,2)$ & $A_{\rm ZT}$ & $A_{\rm CP}$ \\
\noalign{\smallskip}\hline\noalign{\smallskip}
(3,2) &  0.44811   & 4.67439  & 17.0885 &  2.11579 & - \\
(5,2) &  0.33255   & 3.77888 & 24.7394 & 2.50355& -\\
(7,2) &  0.25756   & 3.48742  & 33.0028 & 2.94199 & -\\
(9,2) &  0.21337  &  3.31052  & 41.4243 & 3.35704 & 3.37005\\
(11,2)   & 0.18197   & 3.19781 & 49.8699 & 3.77325   & 3.77022 \\ 
(13,2) & 0.15854   &  3.11985 & 58.3302 & 4.19026  & 4.16099\\
(15,2)   & 0.14041   & 3.06279  & 66.8002  & 4.60785  & 4.54471 \\
(17,2)  & 0.12598   & 3.01920  & 75.2768  & 5.02576 & 4.92303 \\ 
(19,2)   & 0.11423  &  2.98489  & 83.7581 & 5.44404  & 5.29699 \\
(21,2)  & 0.10448      & 2.95711  & 92.2429 & 5.86244  & 5.66758 \\
(23,2)  & 0.09625      & 2.93424  & 100.730 & 6.28113  & 6.03532 \\
(25,2)  & 0.08923    & 2.91503  & 109.220 & 6.69984  & 6.40082 \\
(27,2) & 0.08315  & 2.89868 & 117.711 & 7.11867  & 6.76416 \\
(29,2) &  0.07785   & 2.88466 & 126.204 & 7.53775  & 7.12593  \\ 
\noalign{\smallskip}\hline
\end{tabular}
\end{table}

\begin{table}[h]
\caption{$T(p,1)$ torus slinky found by optimising volume fraction. First column is the knot type, following columns are geometric quantities that characterize the knot; $a/R$ and $H/D$. The fourth column is the rope length, $L(p,1)$ of the knot computed using the criterion of maximal $f_V$. The last column is the aspect ratio, $A$, of the knot.}
\label{tab:2}       
\begin{tabular}{ccccc}
\hline\noalign{\smallskip}
Type  & $a/R$  & $H/D$  & $L(p,1)$ & $A_{\rm CP}$ \\
\noalign{\smallskip}\hline\noalign{\smallskip}
(9,1)   & 0.16968   &1.35594  &22.3099  &3.3413 \\ 
(11,1)   & 0.14972    &1.31317  & 27.8621  &3.72327    \\ 
(13,1)   & 0.13481     &1.28131  &33.6390  &4.09197   \\ 
(15,1)   & 0.12301     &1.25664  &39.5856  &4.45214 \\ 
(17,1)   & 0.11330    &1.23694  &45.6612  &4.80656 \\ 
(19,1)   & 0.10512    &1.22084  &51.8356  &5.15696  \\ 
(21,1)   & 0.09811    &1.20741  &58.0886  &5.50436  \\ 
(23,1)   & 0.09200    &1.19604  &64.4035  &5.84955  \\ 
(25,1)   & 0.08664   &1.18628  &70.7694  &6.19298  \\ 
(27,1)   &  0.08188   &  1.17780 &  77.1777 &  6.53499 \\ 
(29,1)   &  0.07763&   1.17036  &  83.6217 &  6.87585  \\ 
\noalign{\smallskip}\hline
\end{tabular}
\end{table}

\begin{table}[h]
\caption{Biological torus molecules: Listed is PDB ID, bibliographical reference, type and the ($p$-fold) symmetry. The last column is the measured aspect ratio, $A$, of the torus molecule using all-atom van der Waals models.}
\label{tab:3}       
\begin{tabular}{ccccc}
\hline\noalign{\smallskip}
PDB ID  & Ref. & Type  & $p$  & $A$\\
\noalign{\smallskip}\hline\noalign{\smallskip}
4H5P &  \cite{raymond2012} & virus & 4 & 2.60 \\
4H5O & \cite{raymond2012} & virus & 5 & 3.28 \\
4H5Q & \cite{raymond2012} & virus & 6 & 2.96 \\
1RXM &  \cite{chapados2004}  & complex  & 6  &  2.71   \\
1I5L &  \cite{toro2001} &  complex  & 7  &  2.41    \\
2C7D & \cite{ranson2006} & chaperone & 7 & 3.33 \\
3KFE & \cite{pereira2010} &  chaperone & 8 & 3.60 \\
1C9S &  \cite{antson1999} & complex & 11 & 4.24 \\
1H5W &  \cite{guasch2002}  & virus  & 12  &  4.24   \\
3KML & \cite{dedeo2010} & virus & 17 & 3.63 \\ 
\noalign{\smallskip}\hline
\end{tabular}
\end{table}

\end{document}